\documentclass[12pt,preprint]{aastex}

\usepackage{rotating}
\shorttitle{H$\alpha$ Expansion of G156.2+5.7}  

\begin{document}

\title{Constraining the Age and Distance of the Galactic Supernova Remnant G156.2+5.7 by H$\alpha$ Expansion Measurements}

\author{
Satoru Katsuda\altaffilmark{1}, 
Masaomi Tanaka\altaffilmark{2},
Tomoki Morokuma\altaffilmark{3},
Robert Fesen\altaffilmark{4},\\
Dan Milisavljevic\altaffilmark{5}
}

\altaffiltext{1}{Department of Physics, Faculty of Science \& Engineering, Chuo University, 1-13-27 Kasuga, Bunkyo, Tokyo 112-8551, Japan}

\altaffiltext{2}{National Astronomical Observatory of Japan, 2-21-1 Osawa, Mitaka, Tokyo 181-8588, Japan}

\altaffiltext{3}{Institute of Astronomy, Graduate School of Science, The University of Tokyo, 2-21-1 Osawa, Mitaka, Tokyo 181-0015, Japan}

\altaffiltext{4}{6127 Wilder Lab, Department of Physics \& Astronomy, Dartmouth College, Hanover, NH 03755, USA}

\altaffiltext{5}{Harvard-Smithsonian Center for Astrophysics, 60 Garden Street, Cambridge, MA 02138, USA}

\begin{abstract}

We present deep H$\alpha$ images of portions of the X-ray bright but optically faint Galactic supernova remnant G156.2+5.7, revealing numerous and delicately thin nonradiative filaments which mark the location of the remnant's forward shock.  These new images show that these filaments have a complex structure not visible on previous lower resolution optical images.  By comparing H$\alpha$ images taken in 2004 at the McDonald Observatory and in 2015--2016 at the Kiso Observatory, we set a stringent 1-$\sigma$ upper limit of expansion to be 0$\farcs06$ \,yr$^{-1}$.  This proper motion, combined with a shock speed of 500\,km\,s$^{-1}$ inferred from X-ray spectral analyses, gives a distance of $\gtrsim$1.7\,kpc.  In addition, a simple comparison of expansion indices of several SNRs allows us to infer the age of the remnant to be a few 10,000\,yr old.  These estimates are more straightforward and reliable than any other previous studies, and clearly rule out a possibility that G156.2+5.7 is physically associated with part of the Taurus-Auriga cloud and dust complex at a distance of 200--300\,pc.  

\end{abstract}
\keywords{ISM: individual (G156.2+5.7) --- shock waves --- supernova remnants}

\section{Introduction}

G156.2+5.7 is a large ($\sim$1.8 degree diameter), Galactic supernova remnant (SNR) discovered by the {\it ROSAT} all-sky survey \citep{1991A&A...246L..28P}.  It is one of the X-ray brightest and radio faintest SNRs known \citep{1992A&A...256..214R,2007A&A...470..969X}.  Because of its strong X-ray flux, G156.2+5.7 has been continuously exploited with X-ray observatories including {\it Ginga} \citep{1993PASJ...45..795Y}, {\it ASCA} \citep{1999PASJ...51...13Y}, {\it RXTE} \citep{2004AdSpR..33..434P}, {\it Suzaku} \citep{2009PASJ...61S.155K,2012PASJ...64...61U}, and {\it XMM-Newton} \citep{2010ASPC..424..179H,2010PASJ...62..219Y}.  However, there are still large uncertainties on some of its basic parameters such as the age and distance.  

All previous X-ray observations have suggested that G156.2+5.7 is a middle-aged (10--30\,kyr) SNR at a distance of 1--3\,kpc, based on the Sedov analysis.  The uncertainties are mainly due to a relatively large range of the estimated ambient densities of 0.01--0.1\,cm$^{-3}$.  From radio observations, \citet{1992A&A...256..214R} found a large HI shell surrounding the remnant in a velocity range from $-41$\,km\,s$^{-1}$ to $-48$\,km\,s$^{-1}$.  If G156.2+5.7 is associated with the HI shell, the distance may be about 7\,kpc, which is a kinematic distance at a radial velocity of $-45$\,km\,s$^{-1}$ \citep{1986PhDT.........9B}.  However, possible non-circular motions of the HI in this region make the kinematic distance uncertain.  Meanwhile, \citet{2007MNRAS.376..929G} detected a number of H$\alpha$ features including faint nonradiative filaments lying along the edge of the X-ray remnant, and relatively bright radiative filaments inside the remnant.  Since some of the bright radiative filaments seem to be coincident with some nearby ($\sim$0.3\,kpc) interstellar clouds, the authors suggested that G156.2+5.7 may be a nearby and young SNR.  In this context, the age and distance of G156.2+5.7 have been a matter of considerable debate.  

One direct way to constrain the age and distance is to measure remnant's expansion, which is a good tracer of the evolutionary state of SNRs.  We present here our new proper-motion measurements of nonradiative filaments, based on the first-epoch observations taken at the McDonald Observatory in early 2004 and our new observations taken at the Kiso Observatory in late 2015 and early 2016.  We find a very slow expansion of $\lesssim 0\farcs06$\,yr$^{-1}$.  The upper limit allows us to estimate the age and distance to be a few 10,000 yr old and $\gtrsim$1.7\,kpc, respectively.  We give information about observations in Section~2.  Analysis and results are presented in Section~3.  Based on these results, we infer the age and distance of G156.2+5.7 in Section~4, and give a conclusion in Section~5.

\section{Observations and Data Reduction} 

\begin{figure}
\begin{center}
\includegraphics[angle=0,scale=1.5]{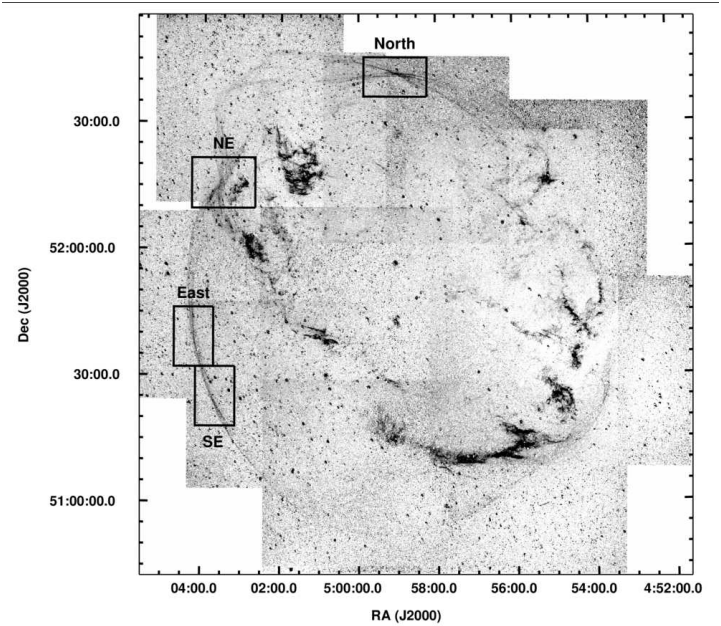}\hspace{1cm}
\caption{H$\alpha$ mosaic image of the G156.2+5.7 SNR taken at the McDonald Observatory \citep{2007MNRAS.376..929G}.  Black boxes indicate locations of the four filament regions shown in Fig.~\ref{fig:image2010}.  
} 
\label{fig:image}
\end{center}
\end{figure}

The first-epoch observations, 13 partially overlapping pointings with each field of view of 46$^{\prime}$ $\times$ 46$^{\prime}$, were performed using the McDonald Observatory 0.76-m telescope in 2004 January.  The data were taken with the H$\alpha$ filter (FWHM of 30 \AA, centered at 6568 \AA) and the adjacent narrow-band filter (FWHM of 30 \AA, centered at 6510 \AA) for the continuum.  Figure~\ref{fig:image} shows a continuum-subtracted H$\alpha$ mosaic of the entire remnant \citep{2007MNRAS.376..929G}.  In measuring proper motions, we use fully-processed, stacked H$\alpha$ images without continuum subtraction, in order to align the images with our images.  Table~\ref{tab:obs} lists basic information for individual first-epoch observations used for our proper-motion measurements \citep[][for more details]{2007MNRAS.376..929G}, as well as the second- and third-epoch observations described below.  

The second-epoch observations were performed on 10 and 11 January and 8 February 2010 using a back-side illuminated 2048 $\times$ 2048 SITe CCD detector attached to the McGraw-Hill 1.3-m telescope at the MDM Observatory at Kitt Peak Arizona.  Although the nights were mostly clear, none were photometric, with seeing varying between 1$\farcs1$ and 2$\farcs5$.  The remnant was imaged using a pair of matched on and off H$\alpha$ interference filters centered at 6568 \AA and 6510 \AA (FWHM = 30 \AA).  Two to four 2000\,s exposures were taken for each of the four target fields.  Morning and evening twilight sky flats along with dark frames were also obtained.  The CCD's 24-$\mu$m size pixels gave an image scale of 0$\farcs508$ and a field of view of approximately 17$^{\prime}$ square.  However, due to the extreme faintness of the remnant's shock filaments, we employed 2 $\times$ 2 pixel on-chip binning to increase the S/N which resulted in a final pixel scale of 1$\farcs015$. This greatly improved the detection of the remnant's optical filaments and matched the telescope's typical image quality of $\sim$1$\farcs5$, while still being better than the 1$\farcs35$ per pixel for the first-epoch H$\alpha$ images of the remnant obtained by \citet{2007MNRAS.376..929G}.  Standard pipeline data reduction of the images was performed using IRAF/STSDAS3. This included debiasing, flat-fielding, and cosmic ray and hot pixel removal. After bias, dark, and flat-field corrections, the H$\alpha$ and continuum images were registered via cross correlation routines in IRAF.

We performed the third-epoch observations of G156.2+5.7, using the Kiso Wide Field Camera \citep[KWFC:][]{2012SPIE.8446E..6LS} mounted on the 105-cm Schmidt telescope at the Kiso Observatory in 2015 November and 2016 January.  The KWFC consists of 8 chips with 2k$\times$4k pixels with a pixel scale of 0$\farcs946$ pixel$^{-1}$, which covers a 2.2 $\times$ 2.2 degree sky area in total.  The large field of view suitably matches the large angular diameter ($\sim$1.8 degree) of G156.2+5.7.  We used an H$\alpha$ narrowband filter (FWHM of 160\AA, centered at 6611\,\AA) and set each exposure time to be 5 minutes with 3$^{\prime}$ dithering.  Overscan subtraction, bias pattern subtraction, and flat fielding with domeflat were performed in a standard manner.  Then, an astrometric solution was obtained with the USNO-B1.0 catalogue using the Optimistic Pattern Matching algorithm \citep{2007PASA...24..189T}, implemented by Dr.\ N.\ Matsunaga for the data reduction for KWFC data.  For each CCD chip, we picked up bad frames suffering from bad seeing, moon light, and cloud occultation.  After rejecting the bad frames, we stacked all frames using SWarp \citep{2002ASPC..281..228B}, for which the effective exposure times are given in Table~\ref{tab:obs}.  The sky background was estimated by relatively large meshes (512 $\times$ 512 pixels or 8$^{\prime}$ $\times$ 8$^{\prime}$) not to oversubtract the filament structures.

\section{Analysis and Results}

\subsection{Filament Morphology}

\begin{figure}
\begin{center}
\includegraphics[angle=0,scale=0.75]{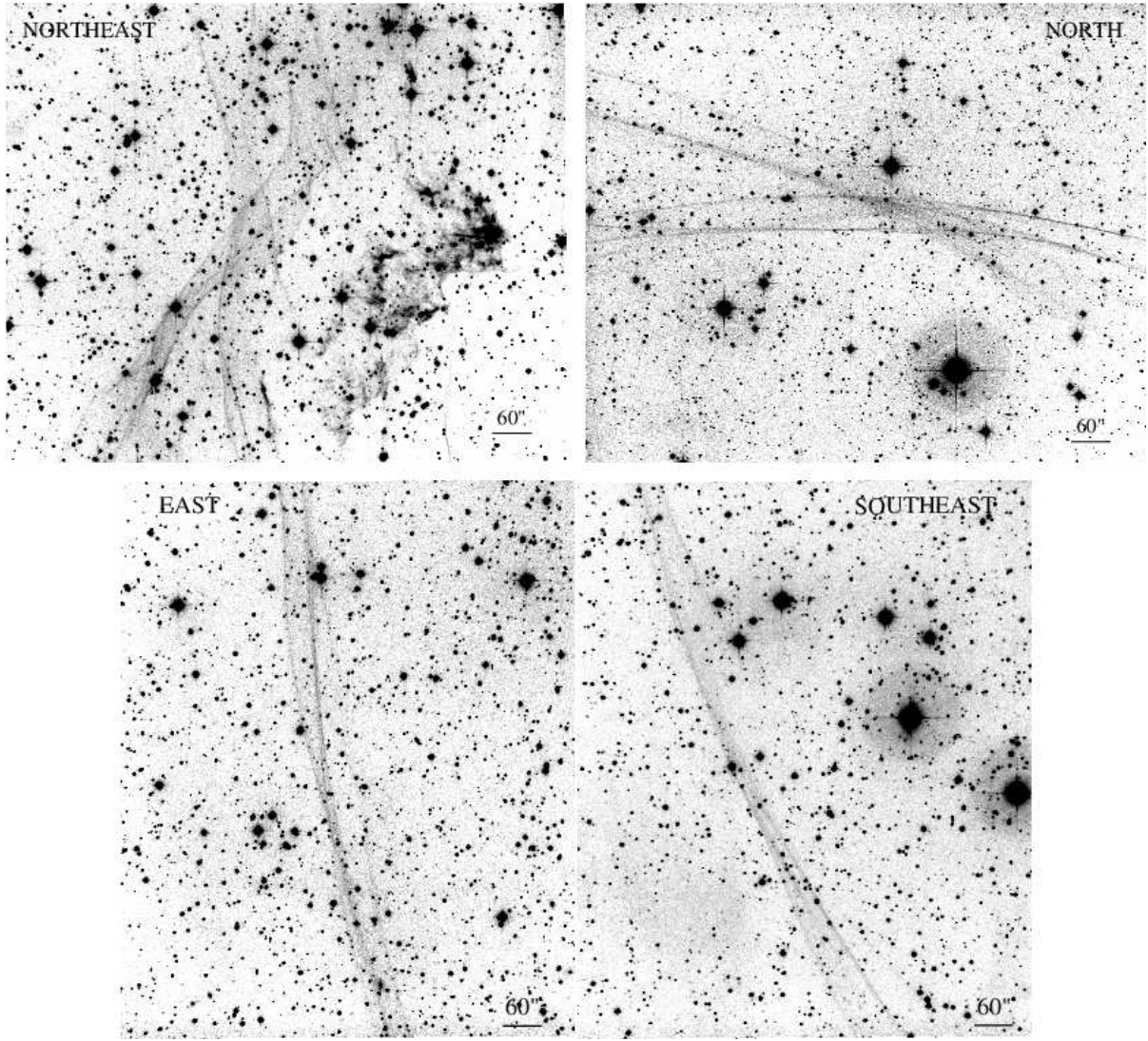}\hspace{1cm}
\caption{H$\alpha$ images of the northeastern (upper left), northern (upper right), eastern (lower left), and southeastern (lower right) portions of G156.2+5.7, taken in 2010 at the MDM Observatory at Kitt Peak Arizona.
} 
\label{fig:image2010}
\end{center}
\end{figure}

We concentrate our imaging program only on the brighter nonradiative filaments primarily found near along the eastern and northern limb regions where multiple thin shock filaments are seen to be closely overlapped in projection.  The complex nature of the remnant's thin nonradiative shock filaments in these two regions is shown in Fig.~\ref{fig:image2010}.  At an resolution of $1\farcs3$ to $2\farcs3$, many of the sharpest filaments are unresolved.  The striking northern criss cross of nonradiative filaments (Fig.~\ref{fig:image2010}), seemingly composed of two long pairs of filaments in the first-epoch images can now be seen to consist of many delicate and overlapping filaments along with faint but noticeable diffuse emission in between filaments.  This complex morphology is similar to that seen in the nonradiative filaments along the northern and eastern limbs of the Cygnus Loop SNR where the shock front is seen edge on and gently warped as it moves through an interstellar medium with small density variations on large scales \citep{1986ApJ...303L..17H,1994ApJ...420..721H,1998ApJS..118..541L,1999AJ....118..942B,2005AJ....129.2268B,2014ApJ...791...30M,2016ApJ...819L..32K}.   

The filamentary emission along the northeastern limb, while qualitatively similar to that seen along the northern limb, differs significantly in the origin of the observed filament arrangement.  The twisted pair of broad filaments are seen to lie eastward out ahead (i.e., at a greater radial distance from the remnant's center) than an adjacent (in projection) complex of radiative shock clouds farther to the west.  The morphology of both the broad nonradiative filaments and the radiative filaments are strikingly similar to that seen in shocked cloud models which investigate the shock passage around and past an ISM cloud \citep{1994ApJ...420..213K,2002AJ....124.2118P,2005ApJ...633..240P,2005ApJ...635..355H,2015ApJS..217...24S,2016MNRAS.457.4470P}.  The filament arrangement seen here appears to be a somewhat more evolved version of a similar shock/cloud encounter involving an isolated ISM cloud located in the southwestern portion of the Cygnus Loop \citep{2002AJ....124.2118P,2005ApJ...633..240P}.  

Emission filaments seen in the two adjacent regions imaged along the eastern and southeastern limb emphasize how complex the remnant's outer optical emission can be, with several thin gently curved and roughly parallel filaments visible along much of the eastern limb.  These images reveal a far richer and more complex filamentary structure than visible in the first-epoch image of the remnant's eastern limb, and nicely coincident with the remnant's X-ray emission in this region.  Portions of some filaments that appear to fade below our image detection limits are presumably due to a decrease in ambient neutral hydrogen density. 

Overall, the remnant's numerous nonradiative filaments seen along the remnant's limbs make G156.2+5.7's optical emission somewhat unusual.  Although the Cygnus Loop exhibits numerous nonradiative filaments along much of its eastern and northern rims, G156.2+5.7's nonradiative filaments encompass most of the remnant's periphery.  In this way, G156.2+5.7 appears somewhat like that of the recently discovered, high latitude remnant, G70.0-21.5, which exhibits nonradiative filamentation nearly everywhere through its optical structure \citep{2015ApJ...812...37F}.

\subsection{Proper Motion}

\begin{figure}
\begin{center}
\includegraphics[angle=0,scale=0.75]{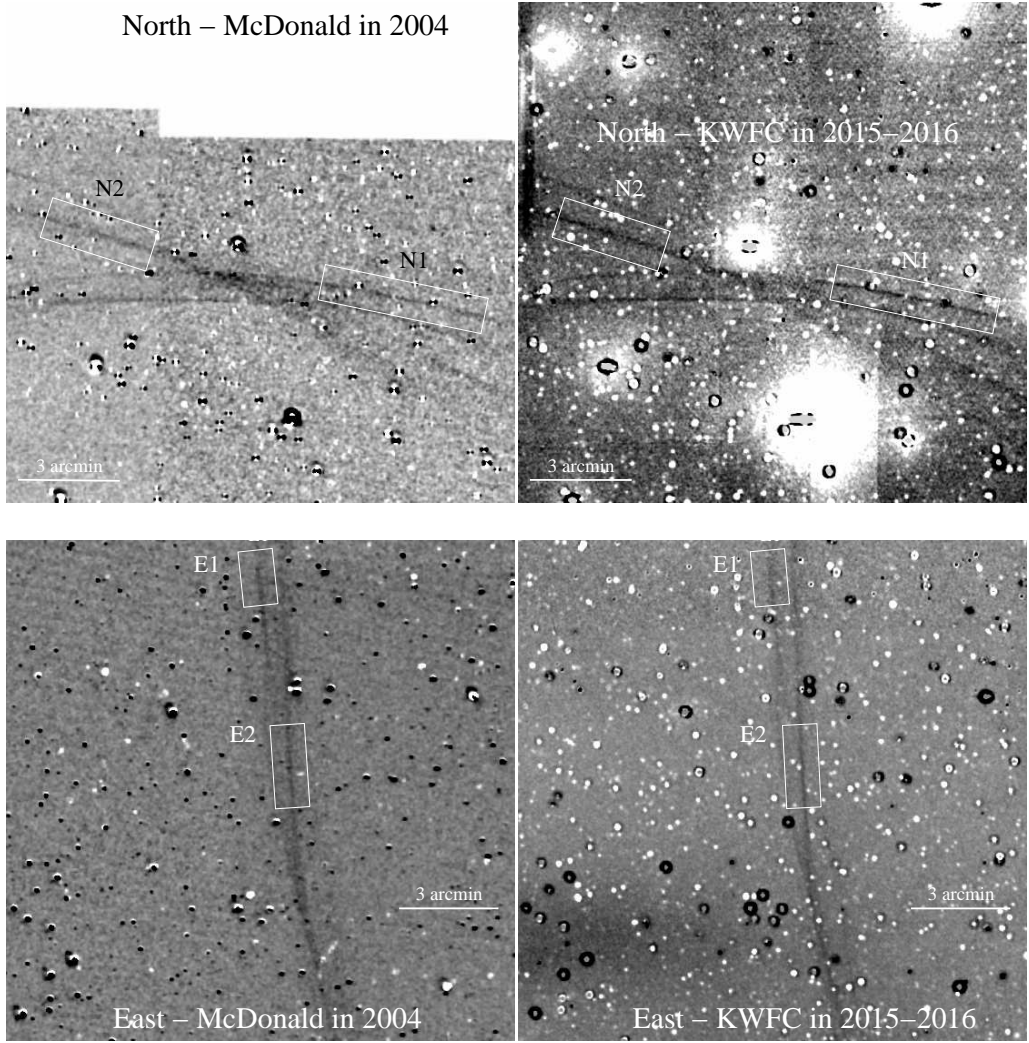}\hspace{1cm}
\caption{Close-up view of continuum-subtracted H$\alpha$ images with the McDonald Observatory (left) and the KWFC (right).  The white boxes show areas (N1, N2, E1, and E2) where we extract radial profiles for proper-motion measurements.
} 
\label{fig:image_comp}
\end{center}
\end{figure}

To measure expansion of the remnant, we selected bright and sharp nonradiative filaments in the northern and eastern limbs, which are indicated by white boxes labelled as N1, N2, E1, and E2 in Fig.~\ref{fig:image_comp} showing continuum-subtracted images taken by the McDonald Observatory in 2004 (left) and the KWFC in 2015-2016 (right).  We note that we failed to obtain meaningful constraints on the proper motions for other outermost nonradiative filaments, mainly due to low signal-to-noise for the first-epoch observations at the McDonald Observatory.  For proper-motion measurements, we basically used only the first- and third-epoch images, because they have a long baseline of 12\,yr and have comparable resolutions with each other.  

The four selected filaments, except for the N1, were detected in two different images obtained in the first-epoch, i.e., two of the A--E in Table~\ref{tab:obs}.  We did not combine the two data, but separately measure proper motions from different pairs because the seeings vary with individual images (see Table~\ref{tab:obs}).  The pairs used to measure proper motions are listed in Table~\ref{tab:registration}.  Although the images in 2004 and 2015-2016 were astrometrically aligned with catalog stars, we further aligned the pair images by using point sources close to the selected filaments to reduce the systematic errors.  Table~\ref{tab:registration} summarizes standard deviations in the position of the point sources after the alignments.  We took the deviation in either RA or Dec.\ direction approximately parallel to the shock normal as a registration uncertainty in our proper-motion measurement.  

\begin{figure}
\begin{center}
\vspace{-0.5cm}
\includegraphics[angle=0,scale=0.55]{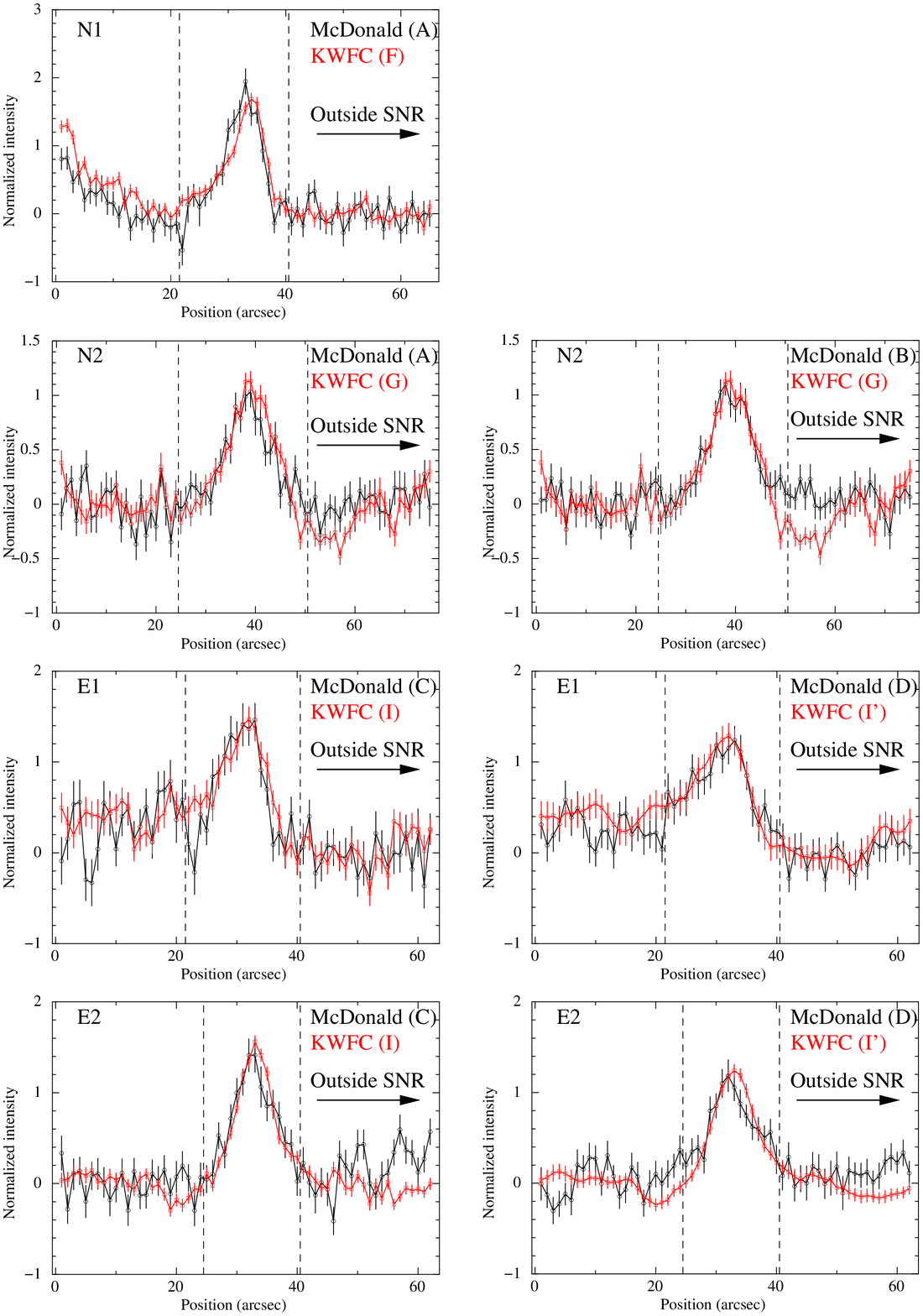}\hspace{1cm}
\caption{H$\alpha$ emission profiles extracted from the four box regions in Fig.~\ref{fig:image_comp}.  Profiles taken at the McDonald Observatory in 2004 are shown in black, and those with the KWFC in 2015 and 2016 are shown in red.  Except for the N1 region, two data sets from the McDonald Observatory are used.  As for the E1 and E2, the KWFC profiles (I') are artificially smoothed to match the image resolution of the McDonald Observatory.  We measure the shifts by using data points within the vertical dashed lines.  
} 
\label{fig:prof}
\end{center}
\end{figure}

To quantitatively measure proper motions, we generated one-dimensional emission profiles across the filament with 1$^{\prime\prime}$ steps.  When generating the profiles, we excluded obvious stars (selected by eyes) in the box regions in order to emphasize the filaments themselves.  For each profile, we subtracted background emission estimated at an off-filament region in the same box area, and then rescaled the height so that the filaments' intensities were equalized between the two epochs.  Thus-derived profiles are presented in Fig.~\ref{fig:prof}, for which the errors, dominated by the noise due to sky background, represent standard deviations of pixel values collected in each bin.  

It should be noted that the seeings for the data sets A and D are relatively worse than those for the other data.  The two data sets were used to measure N1 (A-F), N2 (A-G), and E1/2 (D-I) as shown in Table~\ref{tab:registration}.  As for the D-I pair, we smoothed the KWFC image, i.e., I which is renamed I' after smoothing, with a Gaussian kernel of 1 $\sigma$ of 1$\farcs5$ to artificially match the resolution between the two images.  On the other hand, as for the A-F and A-G pairs as well as all of the other pairs, we did not smooth the KWFC images because the effective resolution (in the direction along the shock normal) is similar between A and F; A suffers from an elongation in east-west direction but not in north-south direction.

Figure~\ref{fig:prof} shows little evidence for shifts of the filaments, which immediately tells us that the expansion is very small.  We quantitatively measured the shifts between the two profiles, following the method established in our previous X-ray proper-motion measurements for other SNRs \citep[e.g.,][]{2008ApJ...678L..35K}.  Briefly, we minimized the $\chi^2$ value for the difference between the third-epoch (KWFC) profile and the shifted first-epoch (McDonald) one, as a function of the amount of the shift.  In calculating the $\chi^2$ values, we used data points within the vertical dashed lines indicated in each panel of Fig.~\ref{fig:prof}, so that we were able to concentrate on the filaments of interest.  The resultant proper motions are summarized in Table~\ref{tab:results}.  The error-weighted mean of their proper motions is calculated to be 0$\farcs{031}\pm0\farcs{010}\pm0\farcs{017}$ yr$^{-1}$, where the first- and second-term errors are responsible for the 1-$\sigma$ statistical and registration uncertainties, respectively.  Therefore, we can constrain the expansion of G156.2+5.7 to be $\lesssim$0$\farcs{06}$ yr$^{-1}$.  We note that the proper motion from the second- (MDM) and third-epoch (KWFC) observations for the E2 filament, which is the best filament to measure a proper motion between the two epochs, is fully consistent with the upper limit derived above, and cannot help reduce the upper limit.

\section{Discussion}

Thanks to a long baseline (12\,yr) between the first-epoch observations at the McDonald Observatory in 2004 and the third-epoch observations at the KWFC Observatory in 2015--2016, we have obtained a stringent upper limit of G156.2+5.7's expansion to be 0$\farcs{06}$ yr$^{-1}$.  Below, we will use the upper limit to constrain the debated age and distance of this remnant.  

\begin{figure}
\begin{center}
\includegraphics[angle=0,scale=0.5]{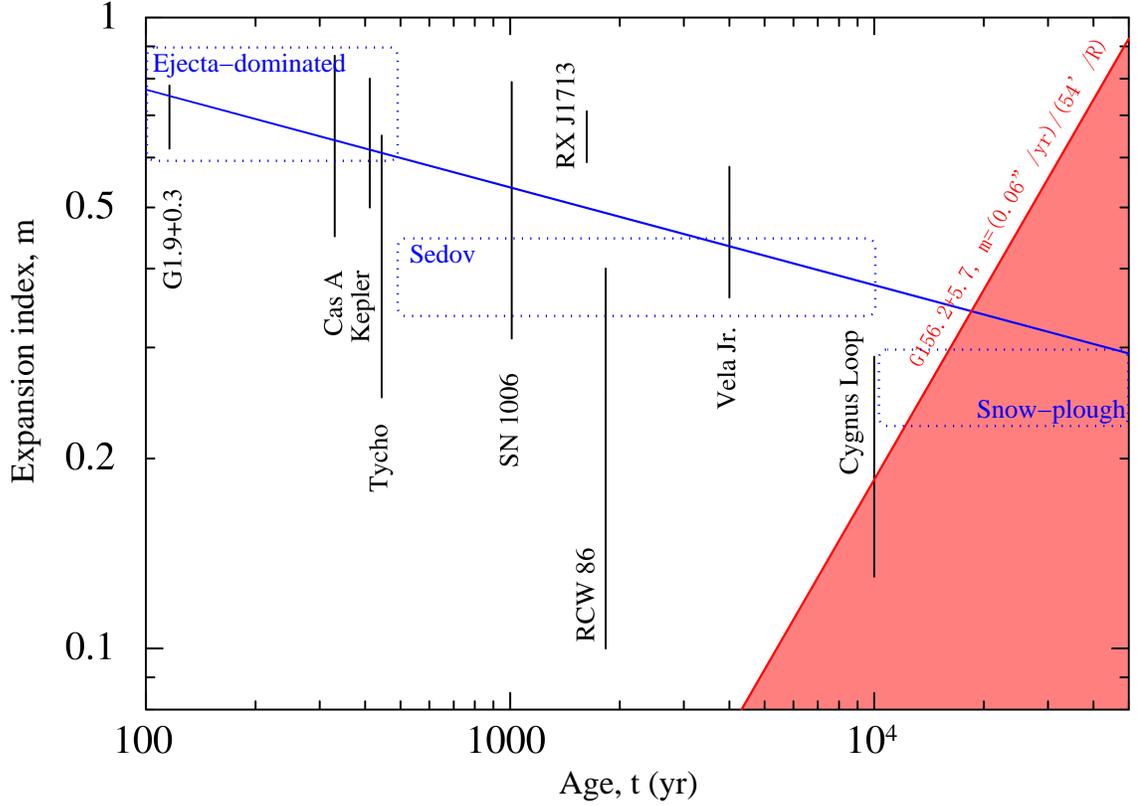}\hspace{1cm}
\caption{Expansion indices against the ages for several SNRs.  We take a range of the expansion indices measured at different portions of each SNR from the literature.  The best-fit power-law model is presented as a blue line, which roughly agrees with $m$-values expected for characteristic evolutionary phases which are indicated by dotted blue areas.  The red area indicates the allowed region for G156.2+5.7, which is limited by the upper limit of the 1-$\sigma$ uncertainty of the expansion.  
} 
\label{fig:expindex_vs_age}
\end{center}
\end{figure}

Our age estimate is simply based on the comparison (extrapolation) of expansion indices ($m$) of some SNRs.  The value of $m$ is defined such that the time evolution of the remnant radius is $R \propto t^{m}$, and is calculated from the proper motion; $m = d({\rm log} R)/d({\rm log} t) = (dR/dt)/(R/t) = \mu t/\theta$, where $R$ is the angular radius to the filament whose proper motion $\mu$ has been measured, $t$ is the age of the remnant, and $\theta$ is the angular radius of the remnant.  In Fig.~\ref{fig:expindex_vs_age}, we plot $m$-values for G1.9+0.3 \citep{2011ApJ...737L..22C}, Cas~A \citep{2003ApJ...589..818D,2009ApJ...697..535P}, Kepler's SNR \citep{2008ApJ...689..225K,2008ApJ...689..231V,2016ApJ...817...36S}, Tycho's SNR \citep{1997ApJ...491..816R,2010ApJ...709.1387K}, SN~1006 \citep{2014ApJ...781...65W}, RX~J1713.7-3946\footnote{In our calculation of the expansion index, we assume that RX~J1713.7-3946 is the remnant of SN~393, which has been a matter of debate \citep{2012AJ....143...27F}.} (Acero et al.\ in prep.; Tsuji et al.\ in prep.), RCW~86 \citep{2013MNRAS.435..910H,2016arXiv160208551Y}, Vela Jr.\ \citep{2008ApJ...678L..35K,2015ApJ...798...82A}, and the Cygnus Loop \citep{2005AJ....129.2268B,2009ApJ...702..327S}, where we take fairly large ranges of the $m$-values from the literature.  

As for G156.2+5.7, we can draw a line on the $m-t$ plane (i.e., a red line in Fig.~\ref{fig:expindex_vs_age}), based on the upper limit on the proper motion and the angular radius.  The $m$-values can be roughly modeled by a power-law function, which is shown as a blue line in Fig.~\ref{fig:expindex_vs_age}.  In fact, the best-fit model roughly agrees with the ranges of $m$-values responsible for characteristic evolutional states indicated by blue dotted areas in Fig.~\ref{fig:expindex_vs_age}: 0.6--0.9 (depending on the density profiles of the ejecta and the circumstellar medium) for the ejecta-dominated phase, $\sim$0.4 for the Sedov phase, and $\sim$0.25 for the snow-plough phase \citep{2012A&ARv..20...49V}.  If we extrapolate the best-fit power-law model to the range of G156.2+5.7, we can infer its age to be $\gtrsim2\times$10$^4$\,yr.  

We caution that the extrapolation of the power-law model may not be valid for G156.2+5.7.  For example, if a blastwave had been recently decelerated due to an interaction with a dense medium (e.g., a molecular cloud), we expect a sharp decrease of the $m$-value when the interaction started.  If it is the case, the age inferred from the power-law extrapolation would be over-estimated.  In fact, emission measure profiles in the eastern and northern limbs exhibit some excesses over the simple Sedov solution, indicating that the blastwave recently encountered a denser material than the shock had propagated before \citep{2009PASJ...61S.155K}.  On the other hand, there is no evidence for dense clouds coincident with the edges of the remnant in these limbs \citep{1992A&A...256..214R}.  Also, the nearly circular shape of the remnant indicates no strong deceleration of the shock in most directions, leaving a question whether or not the blastwave encountered a dense medium in the east and north.  In this context, it is important to reveal the expansion globally, especially in the southwestern limb where we see very faint nonradiative filaments.  This is impossible with the current data set, and is left as our future work.  

The distance can be derived from the shock speed divided by the proper motion.  Using a shock speed of $v_{\rm s}\sim$500\,($kT$/0.3\,keV)$^{0.5}$\,km\,s$^{-1}$ \citep{2009PASJ...61S.155K,2012PASJ...64...61U} and a proper motion of $\lesssim$0$\farcs{06}$ yr$^{-1}$, we can derive a distance of $\gtrsim$1.7\,($v_{\rm s}$/500\,km\,s$^{-1}$)\,($\mu$/0$\farcs{06}$ yr$^{-1}$)$^{-1}$\,kpc.  Therefore, we can clearly rule out the possibility that G156.2+5.7 is a nearby SNR --- a possibility proposed based on the assumption that the remnant is associated with the nearby Taurus-Auriga cloud complex at $\sim$0.3\,kpc \citep{2007MNRAS.376..929G}.  

It should be pointed out that such a large distance is supported by the larger intervening material ($N_{\rm H}\sim$3--5$\times$10$^{21}$\,cm$^{-2}$) with {\it ASCA} \citep{1999PASJ...51...13Y} and {\it Suzaku} \citep{2009PASJ...61S.155K,2012PASJ...64...61U} than that initially estimated ($N_{\rm H}\sim$9$\times$10$^{20}$\,cm$^{-2}$) with {\it ROSAT} \citep{1991A&A...246L..28P}.  In fact, the revised absorption is consistent with the total line-of-sight HI column density to the direction of G156.2+5.7, i.e., ($\ell$, $b$) = (156.2, 5.7), of 3.0--3.5$\times$10$^{21}$\,cm$^{-2}$ \citep{2005A&A...440..775K,1990ARA&A..28..215D}.  This would place the SNR at an edge (or even outside) of the Galactic disk, in agreement with a far distance, e.g., $\sim$350\,pc away from the Galactic plane at a distance of 3.5\,kpc.  

We note that the temperature of 0.3\,keV adopted to estimate the shock speed is taken from the electron temperature measured in the the outermost 2$^{\prime}$ region in the northern and eastern directions \citep{2009PASJ...61S.155K}.  This implicitly assumes electron-ion temperature equilibration, which may not be the case for the filaments of our interest.  If equilibration has not yet been established, the shock would be faster than 500\,km\,s$^{-1}$, as discussed in \citet{2012PASJ...64...61U}, leading to a larger distance to the SNR.  On the other hand, the electron temperature immediately behind the shock may be smaller than 0.3\,keV, given the gradual temperature decrease toward the SNR edge \citep{2009PASJ...61S.155K}.  This would make the shock slower, canceling out the effect of temperature non-equilibration to some extent.  In this context, we believe that the shock speed of 500\,km\,s$^{-1}$ is a good approximation at this moment.  However, it is, of course, desired to perform high-resolution spectroscopy of a nonradiative filament to obtain an accurate shock speed and distance.  

\section{Conclusion}

We find that the expansion of the Galactic SNR G156.2+5.7 is relatively slow, with $\lesssim$0$\farcs{06}$ yr$^{-1}$.  Such a slow expansion allows us to estimate the age and distance of this remnant to be a few 10,000 yr and $\gtrsim$1.7\,kpc, respectively.  Our estimates have less model dependencies than any other previous estimates, and rule out the possibility that G156.2+5.7 is a nearby ($\sim$0.3\,kpc) and young (a few 1000 yr) SNR.

\acknowledgments

We thank all the members at the Kiso Obsesrvatory for allowing us to perform our observations and kind technical cares.  The authors acknowledge Dr.\ N.\ Matsunaga for the software to conduct comparisons between our image coordinates and the catalogue values.  This work is supported by Japan Society for the Promotion of Science KAKENHI Grant Numbers 25800119 and 16K17673 (S.\ Katsuda), 15H00788 and 15H02075 (M.\ Tanaka).


\begin{deluxetable}{lcccccccccc}
\tabletypesize{\tiny}
\tablecaption{H$\alpha$ Image Observations of G156.2+5.7}
\tablewidth{0pt}
\tablehead{
\colhead{Date}&\colhead{Telescope / CCD-ID}&\colhead{RA, Dec.\ (ID)}&\colhead{Pixel scale ($^{\prime\prime}$ pixel$^{-1}$)}&\colhead{Filter coverage (\AA)}&\colhead{Exposure (s)}&\colhead{Seeing ($^{\prime\prime}$)}
}
\startdata
2004-01-12 & McDonald 0.76 m & 4:58:30.5, 52:23:31.5 (A) & 1.354 & 6553--6583 & 1000 & 5.6 \\
2004-01-18 & McDonald 0.76 m & 5:01:42.9, 52:24:29.3 (B) & 1.354 & 6553--6583 & 1000 & 4.0 \\
2004-01-18 & McDonald 0.76 m & 5:03:11.5, 51:46:46.2 (C) & 1.354 & 6553--6583 & 1000 & 3.1 \\
2004-01-18 & McDonald 0.76 m & 5:02:01.3, 51:25:26.4 (D) & 1.354 & 6553--6583 & 1000 & 6.0 \\
2004-01-19 & McDonald 0.76 m & 5:02:49.5, 52:33:24.4 (E) & 1.354 & 6553--6583 & 1000 & 4.1 \\
2010-01-11 & McGraw-Hill 1.3 m & 4:59:02.8, 52:39:07.5 & 1.015 & 6553--6583 & 2000 & 1.3 \\
2010-01-12 & McGraw-Hill 1.3 m & 5:03:29.1, 52:14:28.1 & 1.015 & 6553--6583 & 2000 & 2.3 \\
2010-02-08 & McGraw-Hill 1.3 m & 5:04:08.5, 51:39:12.6 & 1.015 & 6553--6583 & 2000 & 1.8 \\
2010-02-08 & McGraw-Hill 1.3 m & 5:03:34.9, 51:25:15.9 & 1.015 & 6553--6583 & 2000 & 1.9 \\
2015-11 \& 2016-01 & KWFC 1.05 m / SITe-3 & 4:56:10.1, 52:38:22.5 (F) & 0.946 & 6518--6681 & 27900 & 3.9 \\
2015-11 \& 2016-01 & KWFC 1.05 m / SITe-4 & 5:03:22.7, 52:38:02.0 (G) & 0.946 & 6518--6681 & 27000 & 3.9 \\
2015-11 \& 2016-01 & KWFC 1.05 m / SITe-2 & 5:03:19.6, 52:04:22.1 (H) & 0.946 & 6518--6681 & 26400 & 4.0 \\
2015-11 \& 2016-01 & KWFC 1.05 m / MIT-4 & 5:03:15.2, 51:30:42.7 (I) & 0.946 & 6518--6681 & 26100 & 3.9 \\
\enddata
\label{tab:obs}
\end{deluxetable}

\begin{deluxetable}{lcccccccccc}
\tabletypesize{\tiny}
\tablecaption{Registration Uncertainties}
\tablewidth{0pt}
\tablehead{
\colhead{Region, Dataset pair}&\colhead{Deviation in RA ($^{\prime\prime}$)}&\colhead{Deviation in Dec.\ ($^{\prime\prime}$)}
}
\startdata
N1, A-F (Pair-1)  & 0.148 & 0.138$^{*}$ \\
N2, A-G (Pair-1)  & 0.215 & 0.185$^{*}$ \\
N2, B-G (Pair-2)  & 0.158 & 0.154$^{*}$ \\
NE, B-H (Pair-1)  & 0.262$^{*}$ & 0.212 \\
NE, E-H (Pair-2)  & 0.150$^{*}$ & 0.183 \\
E1 \& E2, C-I (Pair-1)  & 0.133$^{*}$ & 0.108 \\
E1 \& E2, D-I (Pair-2) & 0.213$^{*}$ & 0.114 \\
\enddata
\tablecomments{$^{*}$We take this value to be a registration uncertainty in our proper-motion measurement.}
\label{tab:registration}
\end{deluxetable}

\begin{deluxetable}{lcccccccccc}
\tabletypesize{\tiny}
\tablecaption{Proper Motions}
\tablewidth{0pt}
\tablehead{
\colhead{Region}&\multicolumn2c{Pair 1}&\multicolumn2c{Pair 2}&\multicolumn2c{Mean}
}
\startdata
 & arcsec & arcsec yr$^{-1}$ & arcsec & arcsec yr$^{-1}$ & arcsec & arcsec yr$^{-1}$ \\
N1 & 0.744$\pm$0.216$\pm$0.138 & 0.062$\pm$0.018$\pm$0.012 & --- & --- & --- & --- \\
N2 & 0.336$\pm$0.336$\pm$0.185 & 0.028$\pm$0.028$\pm$0.015 & -0.192$\pm$0.312$\pm$0.154 & -0.016$\pm$0.026$\pm$0.013 & 0.048$\pm$0.228$\pm$0.120 & 0.004$\pm$0.019$\pm$0.010 \\
NE & 0.12$\pm$0.228$\pm$0.262 & 0.010$\pm$0.019$\pm$0.022 & 0.000$\pm$0.252$\pm$0.150 & 0.000$\pm$0.021$\pm$0.013 & 0.060$\pm$0.168$\pm$0.156 & 0.005$\pm$0.014$\pm$0.013 \\
E1 & 0.516$\pm$0.372$\pm$0.133 & 0.043$\pm$0.031$\pm$0.011 & -0.408$\pm$0.480$\pm$0.213 & -0.034$\pm$0.040$\pm$0.018 & 0.168$\pm$0.300$\pm$0.132 & 0.014$\pm$0.025$\pm$0.011 \\
E2 & 0.396$\pm$0.264$\pm$0.133 & 0.033$\pm$0.022$\pm$0.011 & 0.588$\pm$0.288$\pm$0.213 & 0.049$\pm$0.024$\pm$0.018 & 0.480$\pm$0.192$\pm$0.132 & 0.040$\pm$0.016$\pm$0.011 \\
\enddata
\tablecomments{The first- and second-term errors represent 1-$\sigma$ statistical and registration uncertainties, respectively.}
\label{tab:results}
\end{deluxetable}

\end{document}